# Logistic equation and COVID-19

Efim Pelinovsky[a-c] (pelinovsky@gmail.com), Andrey Kurkin[a] (aakurkin@gmail.com, corresponding author), Oxana Kurkina[a] (oksana.kurkina@mail.ru), Maria Kokoulina[a] (kokoulinamaria97@gmail.com) and Anastasia Epifanova[a] (epifanova.anastasia.s@gmail.com)

[a] Nizhny Novgorod State Technical University n.a. R.E. Alekseev, Minin st., 24, Nizhny Novgorod, 603950, Russia

[b] National Research University – Higher School of Economics, Myasnitskaya st., 20, Moscow, 101000, Russia

[c] Institute of Applied Physics, Nizhny Novgorod, Ul'yanov st., 46, Nizhny Novgorod, 603950, Russia

**Abstract**

The generalized logistic equation is used to interpret the COVID-19 epidemic data in several countries: Austria, Switzerland, the Netherlands, Italy, Turkey and South Korea. The model coefficients are calculated: the growth rate and the expected number of infected people, as well as the exponent indexes in the generalized logistic equation. It is shown that the dependence of the number of the infected people on time is well described on average by the logistic curve (within the framework of a simple or generalized logistic equation) with a determination coefficient exceeding 0.8. At the same time, the dependence of the number of the infected people per day on time has a very uneven character and can be described very roughly by the logistic curve. To describe it, it is necessary to take into account the dependence of the model coefficients on time or on the total number of cases. Variations, for example, of the growth rate can reach 60%. The variability spectra of the coefficients have characteristic peaks at periods of several days, which corresponds to the observed serial intervals. The use of the stochastic logistic equation is proposed to estimate the number of probable peaks in the coronavirus incidence.

1. **Introduction**

Already in this century, several global epidemics have broken out (bovine spongiform encephalopathy, avian influenza, severe acute respiratory syndrome (SARS), etc.). The latest coronavirus epidemic (CODIV-19) struck everyone with its scale and affected literally all countries forced to take emergency measures to prevent the infection spread of (closure of state borders, quarantine, self-isolation, temporary work break of many enterprises and institutions, transition to distance work and training). The number of people infected in the world exceeds 4.89 million people (the data from end-May 2020), and the number of deaths is more than 320,000 people. General information about this viral infection can be found on the Internet. The dynamics of the disease spread is illustrated in Fig. 1, built according to the World Health Organization (WHO) website (https://www.who.int/emergencies/diseases/novel-coronavirus-2019/situation-reports) on 05/20/2020. In this figure, the growth in the number of coronavirus cases in the world and in several countries is indicated in a semi-logarithmic scale. The dashed lines show exponential asymptotics corresponding to doubling the number of cases in a certain number of days. Asterisks indicate the days when countries introduced restrictive measures. As one can see, the nature of the epidemics spread in each country follows almost the same scenario, first there is an exponential growth (or close to exponential) of the number of infected people, and then this growth slows down (however, the numerical values of the constants describing these curves are different for different countries). In some countries, the number of cases is no longer increasing, so the coronavirus epidemic in these countries is almost over. In other countries, the curves in these coordinates are still almost straight lines, which means an exponential increase in the number of cases, and the epidemic has not yet reached its peak. In general, these curves are quite smooth, although some of them show bends associated with the action of certain quarantine measures.

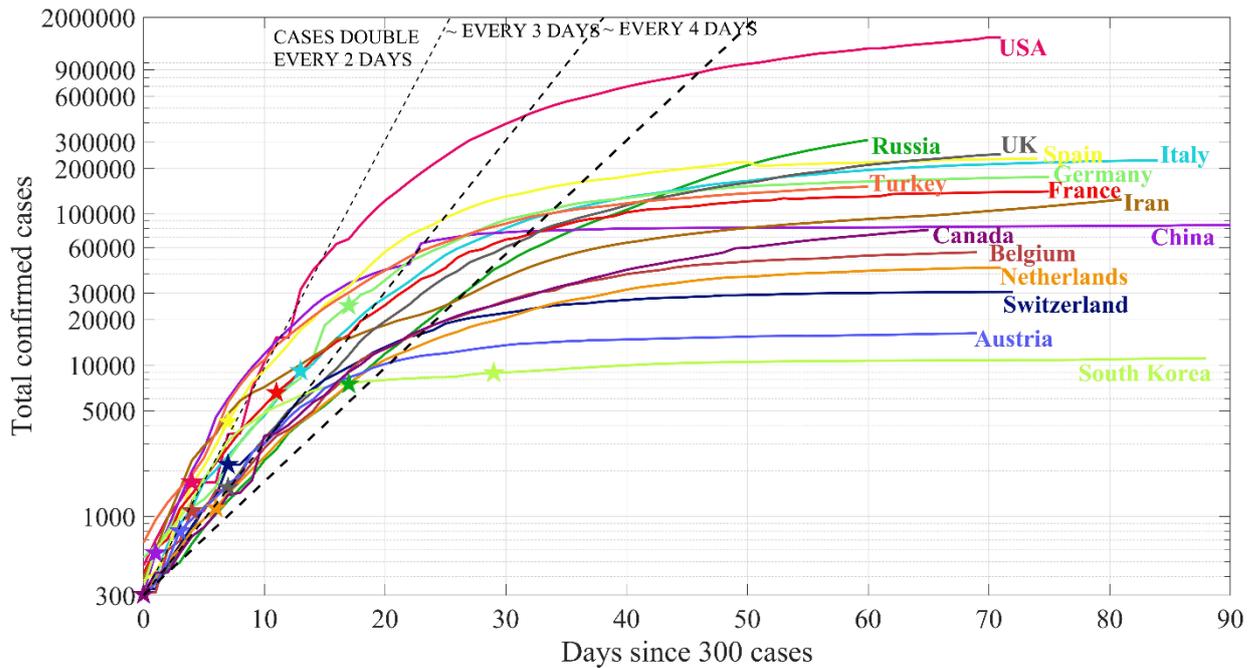

Fig. 1. The confirmed number of people infected with the coronavirus on 05/20/2020.

(Source: WHO data https://www.who.int/emergencies/diseases/novel-coronavirus-2019/situation-reports)

Fig. 2 presents the dynamics of the infection by days, built on the same data. In contrast to Fig. 1, the curves in Fig. 2 are not smooth, and sporadic outbreaks of the number of cases are noticeable in them, which is caused by many, often unpredictable reasons. These data show that in the dynamics of the epidemic spread there are different scales from several months (the total epidemic duration), to several weeks (the incubation period), and even up to several days (the serial interval and local causes). Some of the scales are associated with certain virus properties, others – with the action of the state and local authorities that introduced restrictive rules. The noted features of the dynamics of the COVID-19 virus spread can be reproduced in mathematical models.

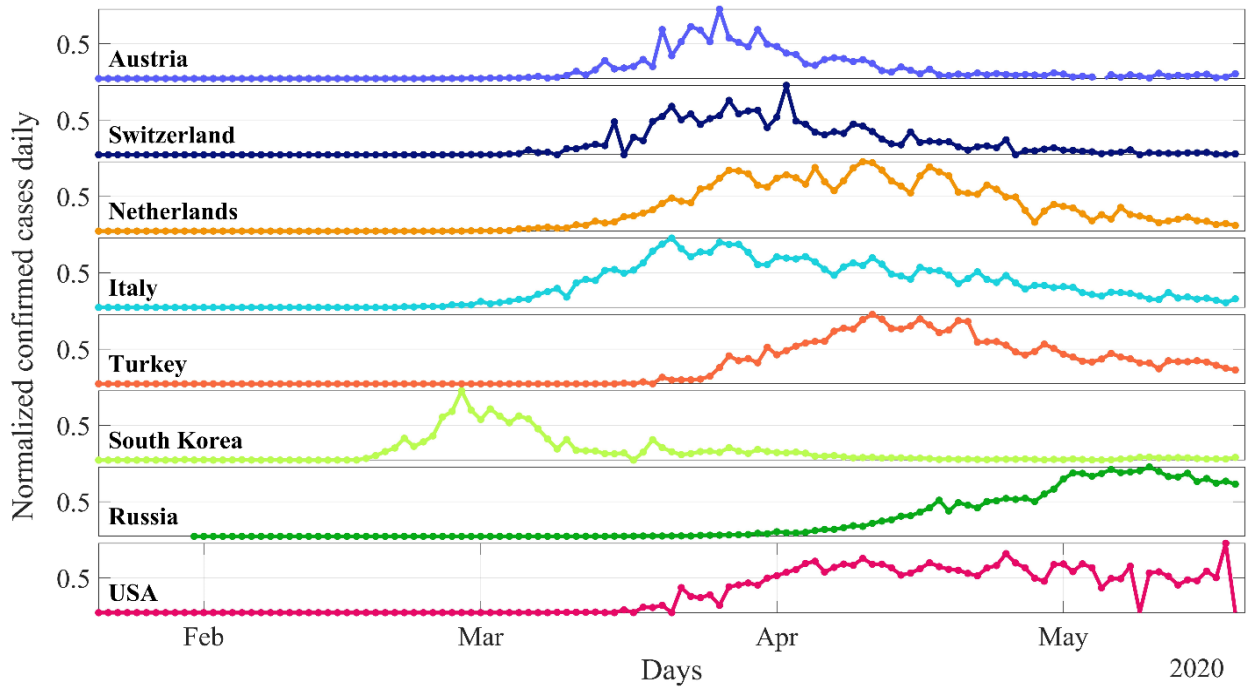

Fig. 2. The number of infected people per day, normalized to the maximum value for each country, according to the same data.

To explain the spread of epidemics and predict their consequences, a number of mathematical models of different complexity levels are used. Historically, the first model is the Verhulst logistic equation [1], representing a nonlinear first-order ordinary differential equation (ODE) with constant coefficients. It is also used as the simplest model to describe the population growth and advertising performance. Qualitatively, it explains the increase in the number of disease cases over the time presented in Fig. 1: the exponential increase in the number of infected people at the initial stage of the epidemic development and the tendency towards a constant value by the end of the epidemic. In the context of COVID-19, this model is used in [2], [3]. The COVID-19 data analysis given in [4], showed that an exponential increase in the number of cases at the initial stage is found mainly in America and Australia, while in many European countries it is a power law. In this case, one can use the generalized logistic equation [5], [6], and it was used in [3], [7], [8], [9]. From the mathematical point of view, the dynamics in the framework of the logistic equation is trivial. More complex dynamics, including chaotic, arise in the different logistic equation or when the delay for the incubation period is accounted for [10], [11], [12], [13], [14], and these models are also used to interpret and forecast COVID-19 [15], [16], [17]. In more complex models, people are divided into different groups: (S) The susceptible class: those individuals who are capable of contracting the disease and becoming infected, (I) The infected class: those individuals who are capable of transmitting the disease to others, and (R) The removed class: infected individuals who are deceased, or have recovered and are either permanently immune or isolated, so the mathematical model called SIR model and its generalizations, includes a higher-order ODE system. The dynamics of such systems has not yet been sufficiently studied, and stochastic oscillations are possible in it [18], [19], [20], [21], [22], [23], [24], [25], [26]. However, models of this level can be comparatively easily implemented, they have shown their effectiveness and are actively used to model the distribution of COVID-19 [27], [28], [29], [30], [31], [32], [33], [34], [35], [36], [37], [38].

There are also models that take into account, for example the super-spreading phenomenon of some individuals or quarantine measures, including social distancing and isolation policies, border control, and a high number in the percentage of reported cases [39 this issue], [40 this issue], [41 this issue].

The statistical methods to forecast the epidemic development, based on Poisson statistics, are also worth mentioning [42], [43], [44], [45].

The main difficulty in applying mathematical models is associated with the uncertainty of the choice of coefficients in the equations. The more complex is the model, the larger is the number of its coefficients. The experience of using models to interpret "old" epidemics may not always help, since the intensity of the virus impact on living organisms changes, many epidemics were local, and, accordingly, measures to prevent the epidemic spread were different. The pattern of the curves shown in Figs. 1 and 2, shows their strong differences for different countries, which is associated with different population density, differences in their customs, traditions and administrative preventive measures. Therefore, any forecasts at the initial stage of the epidemic development regarding its final stage are very rough and unreliable. As the epidemic develops, more and more constants in the equations can be determined from medical databases, but the previous constants are also corrected. Therefore, in essence, for prognostic purposes, equations with variable coefficients are solved, which mathematical properties (existence, convergence and stability) are not defined. As a result, different models with permanently "corrected" coefficients can lead to close forecast results for a short time. At the same time, for long-term forecasts, it is necessary to understand the possible temporal variability of the model coefficients, and their influence on the character of the obtained solutions.

In this study, we will try to assess the character of the scatter of the logistic model coefficients and its generalizations on the basis of the currently available COVID-19 data. The data of the epidemic development were used for the following countries: Austria, Switzerland, the Netherlands, Italy, Turkey and South Korea. Section 2 presents the classical logistic equation and shows the calculations of the coefficient average values within this equation for the above mentioned countries. It has been shown that this model with a high determination coefficient is suitable to describe the number of patients with coronavirus in most countries, except for South Korea. To take into account the data randomness on the number of cases per day, it is proposed to switch to a stochastic logistic equation with external force. The spectral and statistical properties of random parameters of this equation are investigated. Section 3 describes the same procedure within the framework of the generalized logistic equation. It is shown that, on average, this model is suitable for all the countries listed above with a high determination coefficient. Section 4 summarizes the results.

## 2. Logistic equation

Here we will give briefly the main information on the logistic equation theory written in the standard ODE form

$$\frac{dN}{dt} = rN\left(1 - \frac{N}{N_\infty}\right). \tag{1}$$

where $N(t)$ is the total number of people affected by the epidemic, $N_\infty$ is the maximum number of the infected people during the whole epidemic, and $r$ is the growth rate of the epidemic. The solution of this equation with constant coefficients can be easily found in the form

$$N(t) = \frac{N_0 N_\infty \exp(rt)}{N_\infty + N_0 [\exp(rt) - 1]}, \tag{2}$$

where $N_0$ is the initial number of the infected people and $t$ is the time from the beginning of the epidemic. At the initial stage of the epidemic, it can be represented by an exponential function

$$N(t) = N_0 \exp(rt), \tag{3}$$

and, if this curve approximates the increase in the number of cases at the initial stage well, we will be able to determine the growth rate $r$. At the same stage, the logistic model can be rejected if the data do not fit in with the exponential dependence. At the same time, the most important characteristic for prediction – the maximum possible number of the infected people $N_\infty$ – can be estimated only at the stage of the noticeable difference between the data and the exponential curve, when the number of sick people is already not small.

To prepare medical institutions to function in an optimal way during an epidemic, it is important to know the number of infected people per day, which is easily obtained when Eq. (2) is differentiated

$$\frac{dN}{dt} = \frac{N_0 N_\infty (N_\infty - N_0) r \exp(rt)}{\left(N_\infty + N_0 [\exp(rt) - 1]\right)^2}, \tag{4}$$

and this curve is nonmonotonic with the maximum given by

$$\max\left(\frac{dN}{dt}\right) = \frac{rN_\infty}{4}, \tag{5}$$

which corresponds to the time (the epidemic peak)

$$T = \frac{1}{r}\ln\frac{N_\infty - N_0}{N_0}. \tag{6}$$

As it can be seen, these characteristics (Eqs. (5) and (6)) can only be estimated when the data are no longer described by an exponential curve and both model parameters $r$ and $N_\infty$ are found or known.

Let us note that the time dependences (2) and (4) are smooth functions, while from Fig. 2 it follows that dependence (4) must be non-smooth and irregular. The study of the resulting irregularity is carried out below.

Since medical statistics operates with the cases per day, it is in fact necessary to solve the difference logistic equation

$$K_n = N_{n+1} - N_n = rN_n\left(1 - \frac{N_n}{N_\infty}\right). \tag{7}$$

After removing the index n, we obtain a simple relationship between the number of cases per day (K) and the total number of cases (N)

$$K = rN\left(1 - \frac{N}{N_\infty}\right), \tag{8}$$

which in these variables is a parabola.

As an example, we will take the data on the coronavirus incidence in several countries where the epidemic is close to its end (at least its active phase is over). These countries are Austria (number of points 58), Switzerland (58 points), the Netherlands (64 points), Italy (72 points), Turkey (49 points) and South Korea (number of points 94). We will operate with the data on the 04/23/2020); they are taken from the WHO data (https://www.who.int/emergencies/diseases/novel-coronavirus-2019/situation-reports). Figure 3 shows the relationship between the number of cases per day (*K*) and the total number of cases (*N*) for each country. Parabolic approximations (the solid lines) arising from (8) are also presented here. Non-simultaneous 95% prediction bounds for response values (the dashed lines) are shown as well.

Evidently, the parabolic approximation of the available data is good enough for almost all of the listed countries ($R^2 > 0.8$), but obviously has low accuracy for South Korea ($R^2 \sim 0.55$). Therefore, later in this section we will not use the data on South Korea, for which the logistic model is not suitable (this case is analyzed in the next section). Despite a good approximation of the data for most countries of the logistic curve, the scatter of points near the parabolic curve is still not small; it indicates that it is necessary to consider the coefficients of the parabolic curve as the time functions, which, in essence, is done in the forecasts when these coefficients are refined when new data appear. Let us, for example, change only the coefficient *r*. Within the framework of the logistic model, this coefficient variability can be determined from the available data by using the following formula, arising from (8):

$$r = \frac{K}{N\left(1 - \dfrac{N}{N_\infty}\right)}. \quad (9)$$

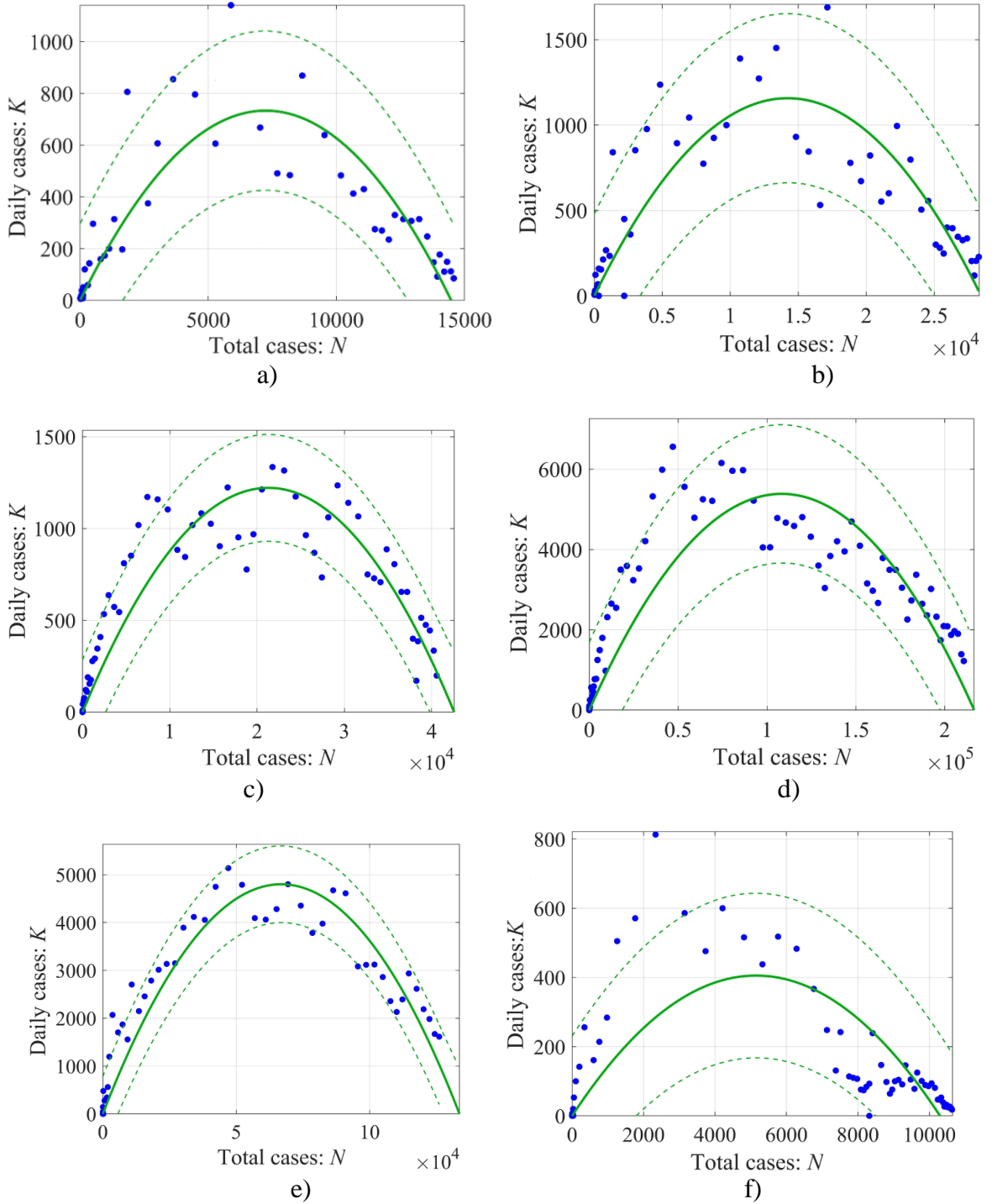

Fig. 3. The relationship between the number of cases per day ($K$) and the total number of cases ($N$). The markers show the data, the solid line is the regression according to Eq. (8), and the dashed lines give non-simultaneous 95% prediction bounds for response values: $a$ – Austria: $N_\infty = 14700$, $r = 0.195$, $R^2 = 0.81$; $b$ – Switzerland: $N_\infty = 28400$, $r = 0.163$, $R^2 = 0.81$; $c$ – The Netherlands: $N_\infty = 42580$, $r = 0.114$, $R^2 = 0.89$; $d$ – Italy: $N_\infty = 216600$, $r = 0.099$, $R^2 = 0.82$; $e$ – Turkey: $N_\infty = 133700$, $r = 0.144$, $R^2 = 0.94$; $f$ – South Korea: $N_\infty = 10300$, $r = 0.158$, $R^2 = 0.55$.

Actually, there are two ways of analyzing this coefficient; it can be either a function of the number of cases (*N*) or time (*t*). In the former case, the logistic equation remains to be the ODE with constant coefficients, but it has a rather complex nonlinearity. In the latter case we come to the ODE with variable coefficients. We study both possibilities of changing the growth rate coefficient *r*. Fig. 4, top, shows the variability of the function *r*(*N*) for Austria and Switzerland. For convenience, we switched to a dimensionless variable $r_{norm} = (r - \langle r \rangle)/\langle r \rangle$, for its variability to be more obvious. Herewith we didn't take into account the first few days, when nothing is clear with the epidemic, and the last few days, when the epidemic was essentially over, since these points correspond to the small values of the denominator in Eq. (9). Reducing the number of points, of course, affects somewhat the average value of this coefficient (for Austria 0.225 instead of 0.195 as in Fig. 3, 0.2 instead of 0.16 for Switzerland), but more important is the demonstration of variability of the coefficient *r*. Functional dependence *r*(*N*) can be rewritten in more familiar terms of temporal variability *r*(*t*), presented in Fig. 4, bottom, where it is demonstrated that this coefficient changes almost every day. As an example, let us give the amplitude spectrum of the coefficient variation *r*, relative to the average, for Austria and Switzerland (Fig. 5, left). Peaks corresponding to intra-weekly variability are clearly visible on the spectrogram due to the fluctuation properties of the epidemic spread, which are different inside condominiums with different apartment numbers or on farms far from each other. In fact, changes in model coefficients can be considered to be random. The probability distribution of the same coefficient for Austria and Switzerland is characterized by the probability density (Fig. 5, right), which is well described by a normal curve. The standard deviation is not small (60-70%), that fact speaks once again about the necessity to take into account the growth rate variability in the epidemic dynamics.

Similar conclusions can be drawn for other countries, but we will not consider them in detail. From the analysis given above it is clear that, on average, the epidemic development in a number of countries is well described by the logistic equation with constant coefficients.

However, to give a more detailed understanding of variations in the number of cases per day, it is reasonable to consider a stochastic logistic equation

$$\frac{dN}{dt} = r(t)N[1 - p(t)N], \quad (10)$$

or its difference analogue, in the general case with two random functions. From the point of view of the available data, the coefficients can be also considered to be random functions of the case number. The properties of a stochastic equation (Eq. (10)) have been investigated yet.

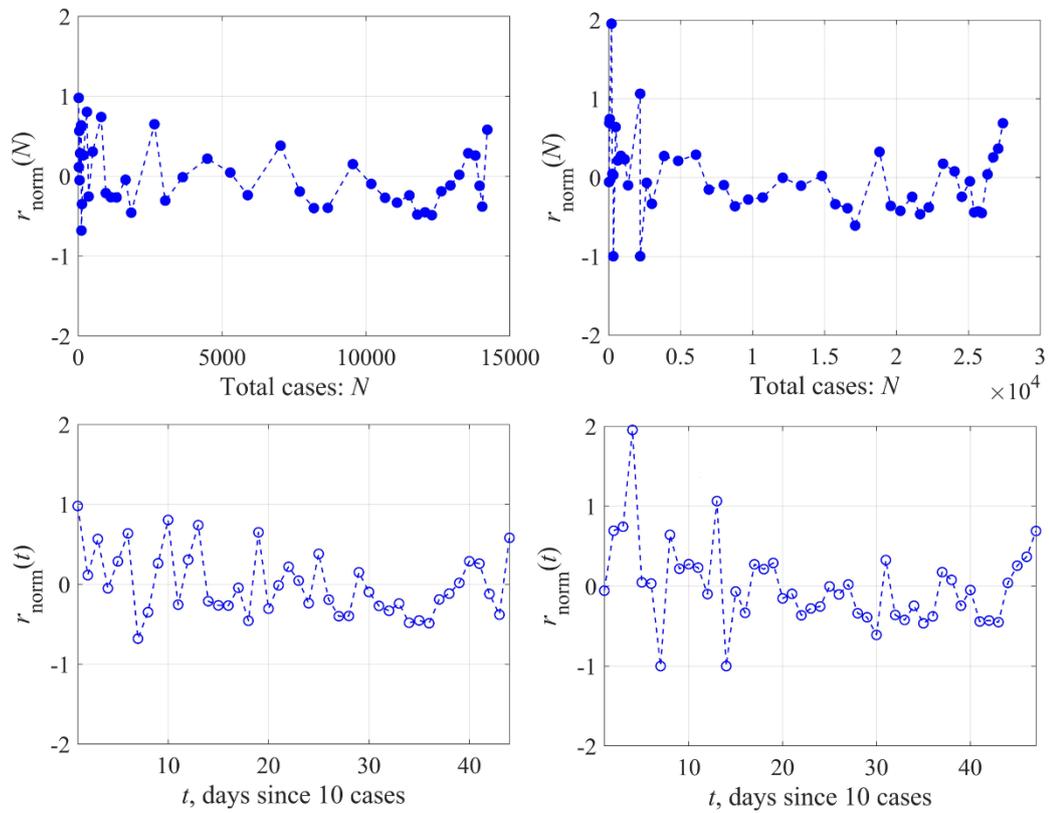

Fig. 4. The normalized growth rate coefficient $r_{norm} = (r - \langle r \rangle)/\langle r \rangle$ as a function of the total number of the infected people ($N$) – the upper panel and of time ($t$) – the lower panel: left – for Austria, right – for Switzerland.

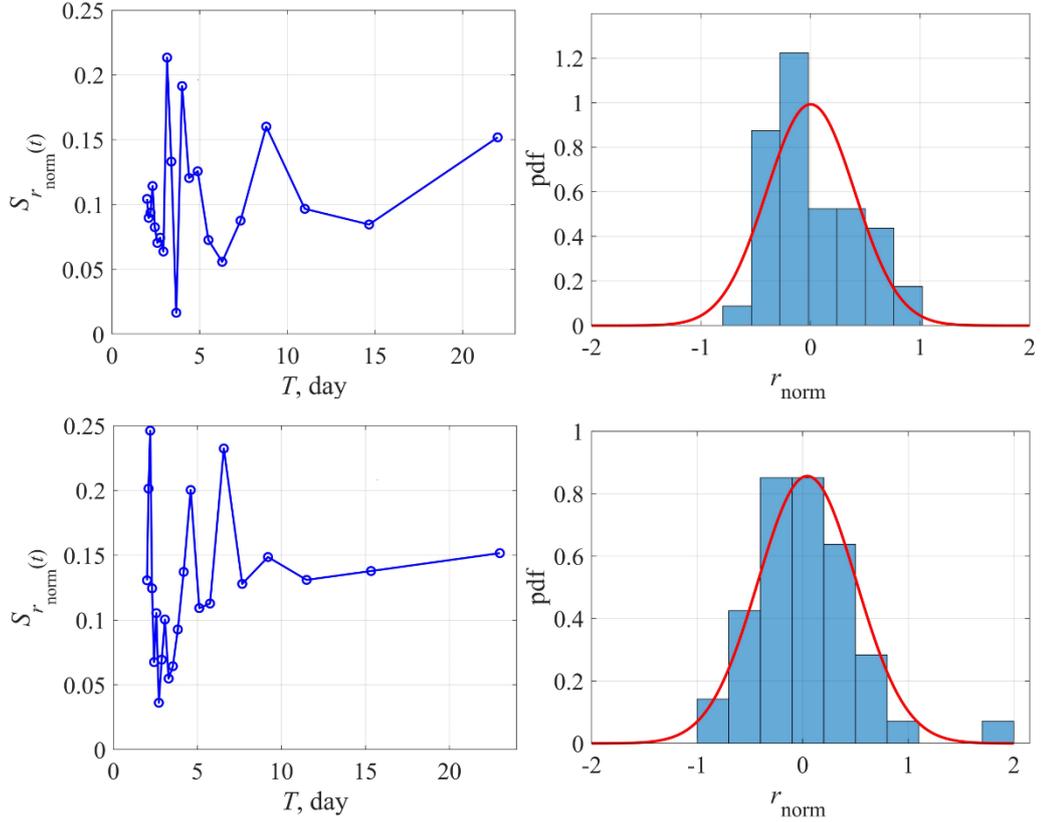

Fig. 5. The spectrum and the distribution histogram for $r(t)$: top – for Austria, bottom – for Switzerland. Distribution parameters for Austria: standard deviation 0.6, skewness 0.7, kurtosis −0.4; for Switzerland: standard deviation 0.7, skewness 1.2, kurtosis −3.2.

In fact, another way to take into account the initial data irregularity is possible, namely, the introduction of an external random force in Eq. (1), as is often done in problems of mechanics:

$$\frac{dN}{dt} = rN\left(1 - \frac{N}{N_\infty}\right) + f(t), \qquad (11)$$

considering all the coefficients constant. The external force is easily found from the available data using the equation following from Eq. (11):

$$f = K - rN\left(1 - \frac{N}{N_\infty}\right). \qquad (12)$$

Fig. 6 shows the dependence of the "external force" $f$ calculated by formula (12) on the total number of cases $N$ (the upper panel) and on the time $t$ (the lower panel) for Austria (left) and Switzerland (right). The magnitude of variations in the external force $f$ is rather small (less than 10% of the total number of cases), and in this sense it does not significantly affect the curve behavior $N(t)$, but it becomes important to analyze the variability of $K(t)$. The probability characteristics of the values of $f$ are shown in Fig. 7 for Austria. The distribution density is well approximated by the Gaussian curve with an average value of 10 and a standard deviation of 132. The differences from the Gaussian curve are characterized by the skewness of 1.1 and kurtosis equal to 2. Similar conclusions can be made for the epidemic data in other countries.

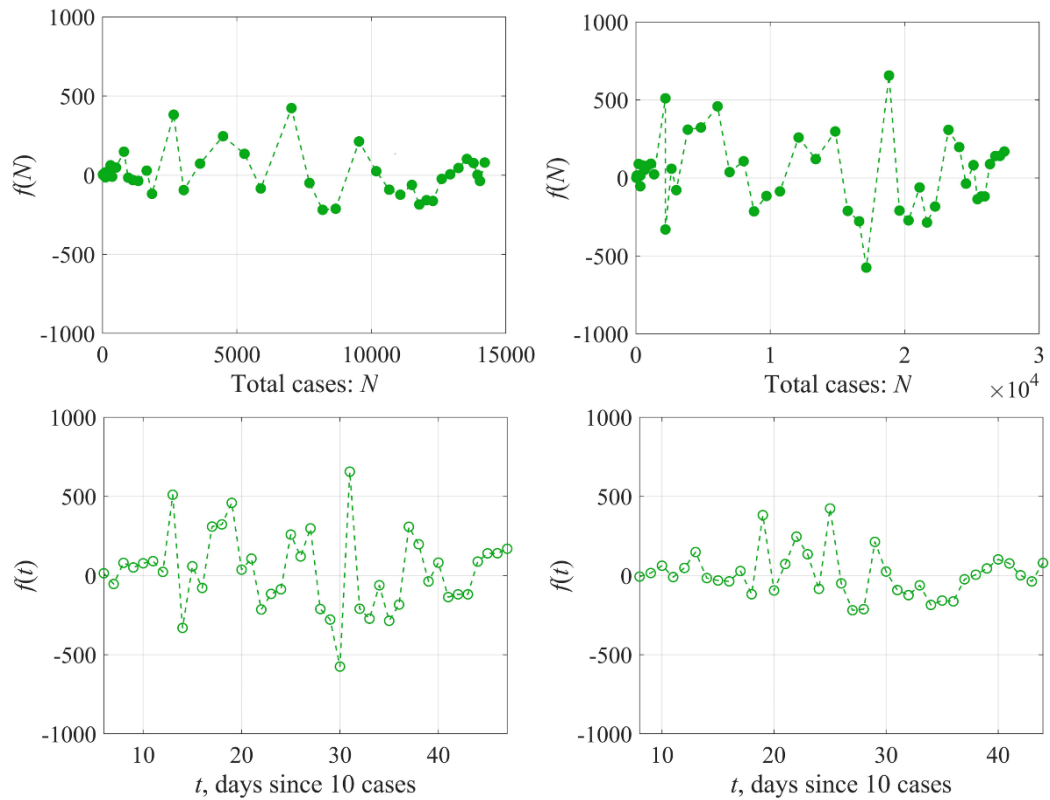

Fig. 6. "External force" $f$ as a function of the total number of cases $N$ on the upper panel and as a function of the time on the lower panel: the left column is for Austria, the right column is for Switzerland.

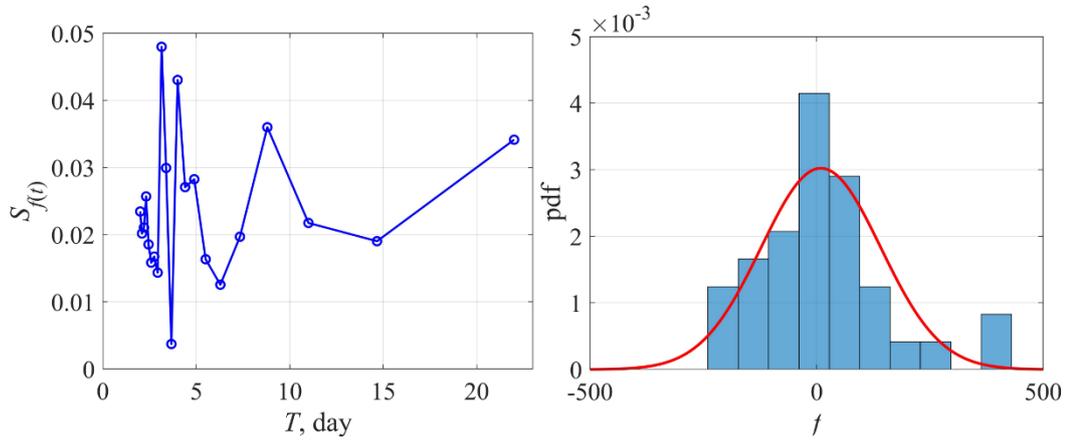

Fig. 7. The spectrum of the "external force" (left) and its probability distribution (right) for Austria. The Gaussian approximation is shown by the red line.

Thus, in principle, stochastic generalizations of the logistic equation can be considered

$$\frac{dN}{dt} = r(t)N[1 - p(t)N] + f(t),  \quad (13)$$

or its difference analogue with an external random force, depending on the number of cases or on time. It will explain the irregularities in the number of cases per day, and the appearance of several peaks of incidence and their duration, which are not predicted by the deterministic logistic equation.

### 3. Generalized logistic model

We will now consider a more general model of a logistic equation containing four constants [5], [6]

$$\frac{dN}{dt} = rN^\alpha \left(1 - \frac{N}{N_\infty}\right)^\beta. \quad (14)$$

When $\alpha = \beta = 1$, Eq.(14) coincides with Eq. (1). Again, our goal is not to solve the equation, but to investigate the relationship between the number of cases per day ($K$) and the total number of the infected people ($N$), which is expressed by the algebraic curve resulting from (14):

$$K = rN^\alpha \left(1 - \frac{N}{N_\infty}\right)^\beta, \qquad (15)$$

Let us consider this model applicability to the description of the development of the COVID-19 epidemic in the same countries as above: Austria, Switzerland, the Netherlands, Italy, Turkey and South Korea. Figure 8 presents the data from medical statistics and approximations from Eq. (15).

It follows from Fig. 8 that the approximation accuracy has increased for all countries ($R^2 = 0.86 - 0.97$), including South Korea ($R^2 = 0.91$), for which a simple logistic model was not suitable at all. In fact, it is the consequence of the general statistics rule that an increase in the number of parameters of the approximation curve leads to an increase in the correlation coefficient, so there is nothing surprising here. More interesting are the magnitude of the exponent indexes in Eqs. (14), (15) and their differences from the unity in a simple logistic model. The power $\alpha$ varies from 0.6 to 1.2, and it is not very far from the unity. Nevertheless, this leads to a qualitative difference in the behavior of the growth curve of the case number at the initial epidemic stage. So, Eq. (14) with a small number of cases is easily solved

$$N = \begin{cases} [(1-\alpha)rt]^{\frac{1}{1-\alpha}}, & 0 < \alpha < 1, \\ N_0 \exp(rt), & \alpha = 1, \\ \dfrac{1}{[(\alpha-1)r(t_0 - t)]^{\frac{1}{\alpha-1}}}, & \alpha > 1 \end{cases} \qquad (16)$$

In contrast to the simple logistic model, where the increase in the case number occurs according to the exponential law, in the generalized logistic model with $\alpha \neq 1$ the growth occurs according to the power law. As it is noted in the Introduction, a power law of gradual increase in the number of cases during the COVID-19 epidemic is typical for many countries [4], so the generalized logistic curve is in good agreement with the medical statistics data.

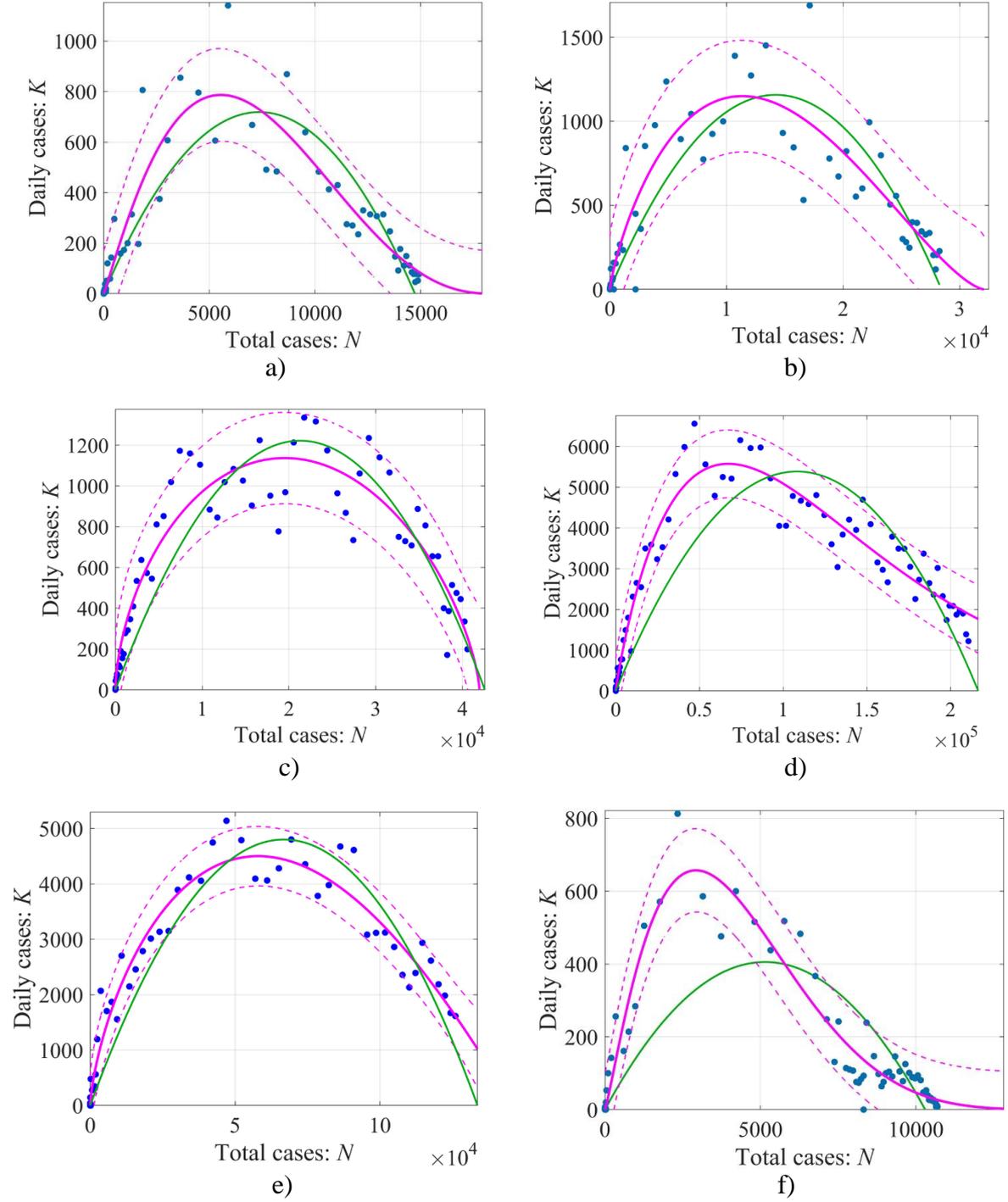

Fig. 8. The relationship between the number of cases per day (K) and the total number of cases (N) in the framework of the generalized logistic model: approximation from Eq. (15) is given by the pink line, the dashed line shows non-simultaneous 95% prediction bounds for response values for the generalized model. The green line shows a parabolic curve in the framework of a simple logistic model (Eq. (8)): $a$ – Austria: $N_\infty = 18500$, $r = 0.151$, $\alpha = 1.1$, $\beta = 2.56$, $R^2 = 0.88$; $b$ – Switzerland: $N_\infty = 32100$, $r = 1.093$, $\alpha = 0.8$, $\beta = 1.5$, $R^2 = 0.86$; $c$ – the Netherlands: $N_\infty = 41950$, $r = 4.702$, $\alpha = 0.6$, $\beta = 0.7$, $R^2 = 0.94$; $d$ – Italy: $N_\infty = 499300$, $r = 0.994$, $\alpha = 0.8$, $\beta = 5.1$, $R^2 = 0.96$; e – Turkey: $N_\infty = 146500$, $r = 3.573$, $\alpha = 0.7$, $\beta = 1.07$, $R^2 = 0.975$; $f$ – Suth Korea: $N_\infty = 16200$, $r = 0.143$, $\alpha = 1.2$, $\beta = 5.4$, $R^2 = 0.91$.

The range of values of $\beta$ is rather wide, from 0.7 to 5.4. The values of $\beta > 1$, lead to an asymmetric deformation of function $K(N)$ towards small values, and means that the epidemic peak is relatively fast, while the end of the epidemic is delayed.

Let us now evaluate the variability of the coefficients of the generalized logistic model from the known data for Austria and Switzerland (to see what differences the generalized logistic model brings) and South Korea, for which a simple logistic model does not work properly. The growth rate variability is given by the formula

$$r = \frac{K}{N^\alpha \left(1 - \frac{N}{N_\infty}\right)^\beta}, \qquad (17)$$

generalizing Eq. (9). Fig. 9 shows the dependence $r_{norm} = (r - \langle r \rangle)/\langle r \rangle$ on the total number of cases ($N$) – left, and on the time ($t$) – right, in the framework of the generalized logistic model (14) for Austria, Switzerland and South Korea (top to bottom). Fig. 10 shows the spectrum of variations and the distribution histogram. For Austria and Switzerland, the characteristics of variability have not changed much compared to those considered above, which is to be expected, since both models give similar results for them. But now we can evaluate the growth rate variability for South Korea, for which a simple logistic model does not work. In general, the characteristics of variability are close for different countries, in the sense that the spectra have peaks at close frequencies and the histograms are qualitatively similar.

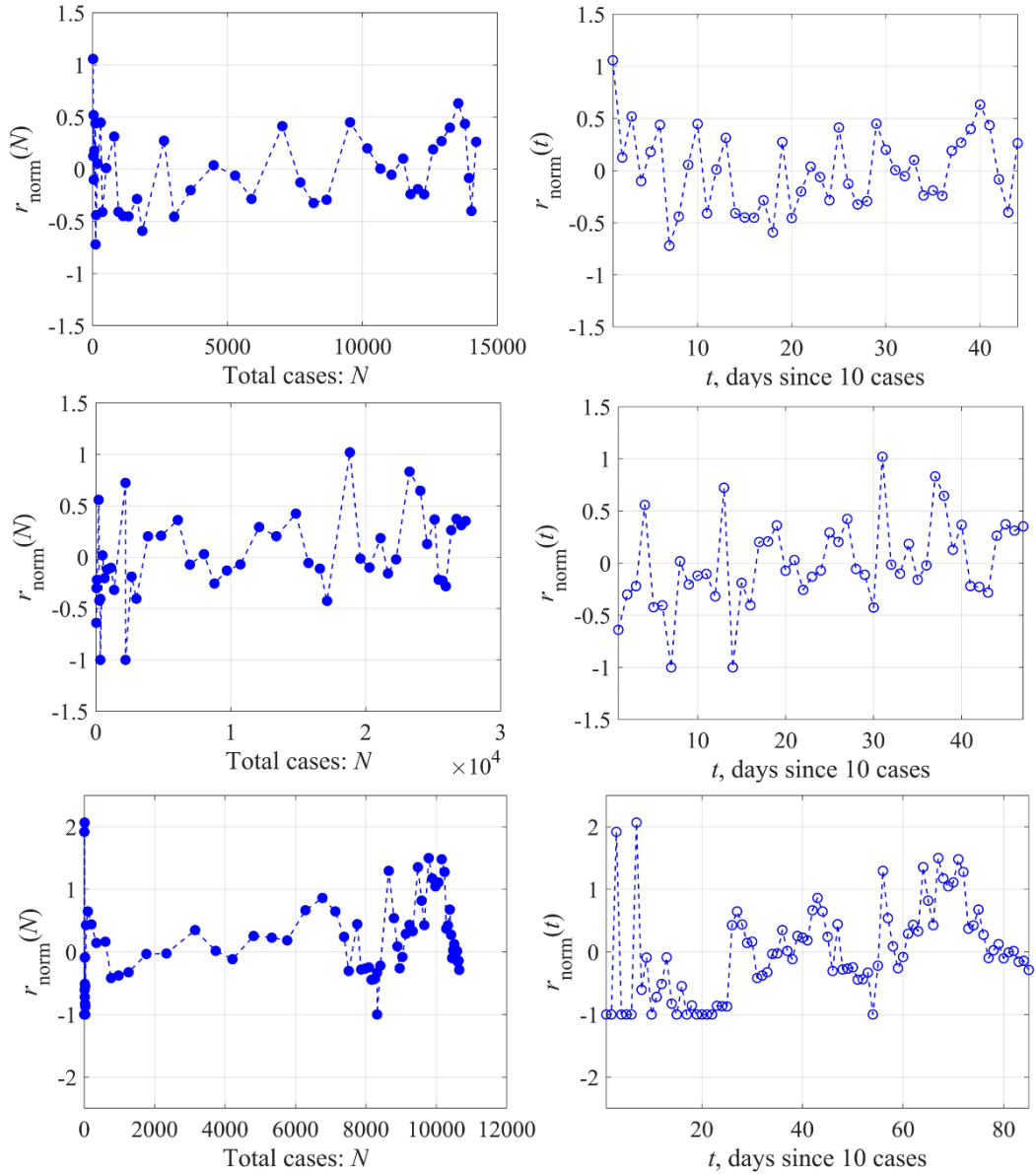

Fig. 9. The normalized growth rate coefficient $r_{norm} = (r - \langle r \rangle)/\langle r \rangle$ as a function of the total cases (left) and of the time (right) in the framework of the generalized logistic model for Austria, Switzerland and South Korea (top to bottom).

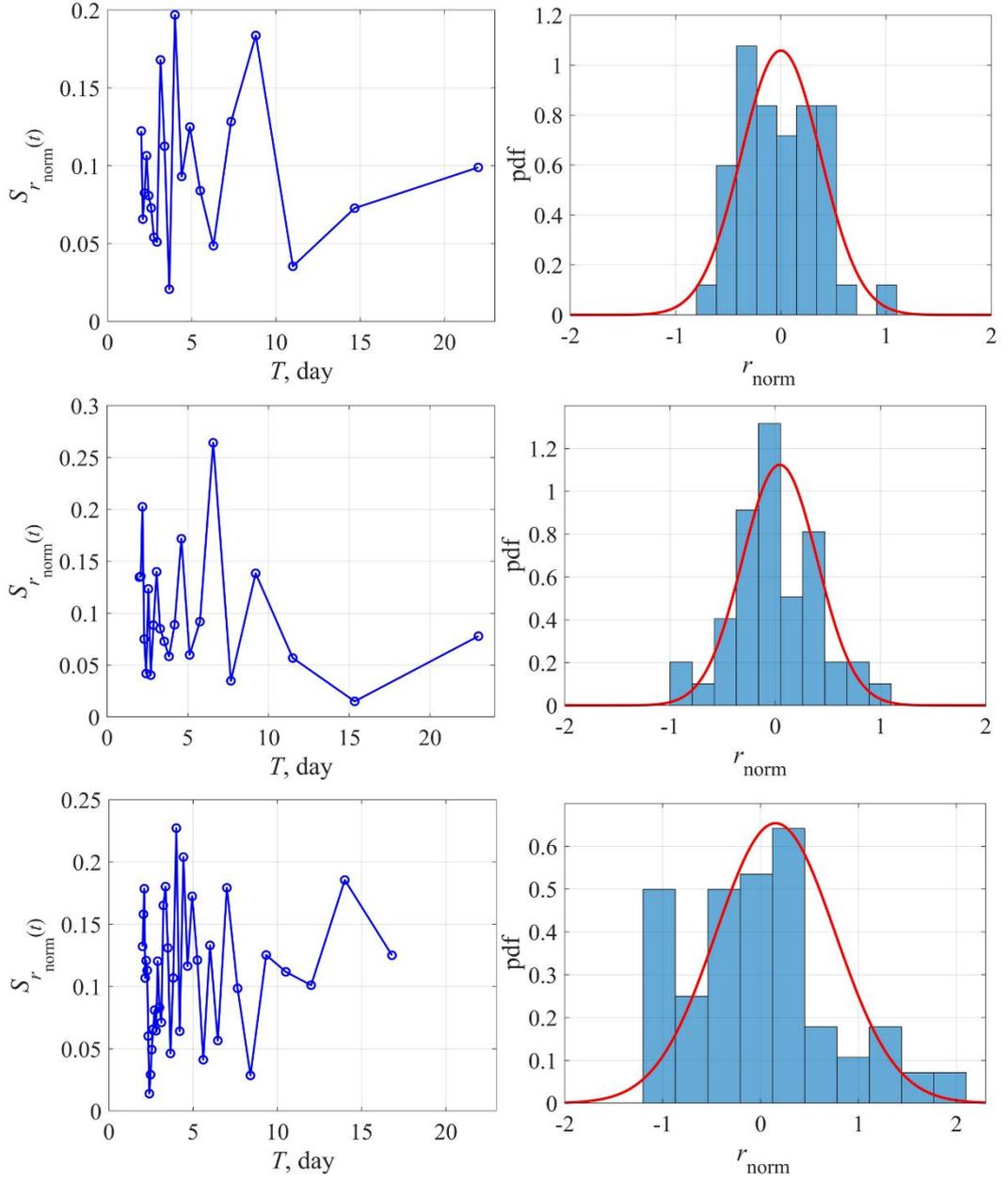

Fig. 10. Spectrum and distribution histogram for *r*(*t*) in the framework of the generalized logistic model for Austria, Switzerland and South Korea (top to bottom). Distribution parameters for Austria: standard deviation 0.6, skewness 0.4, kurtosis –0.17; for Switzerland: standard deviation 0.6, skewness 0.03, kurtosis 0.4; and for South Korea: standard deviation 0.9, skewness 0.6, kurtosis –0.07.

Similarly, we can relate the data discrepancy with the theory using the "external force" introduced analogously to Eq. (12):

$$f = K - rN^\alpha \left(1 - \frac{N}{N_\infty}\right)^\beta. \qquad (18)$$

Fig. 11 shows the calculated dependences of external force on the number of patients and on the time. Its spectral and probability characteristics are illustrated by Fig. 12. The corresponding graphs are similar to those within a simple logistic curve. We would like to emphasize once again that a larger number of countries are properly described by this model, in particular South Korea, demonstrating the qualitatively identical nature of the variability of the logistic model coefficients.

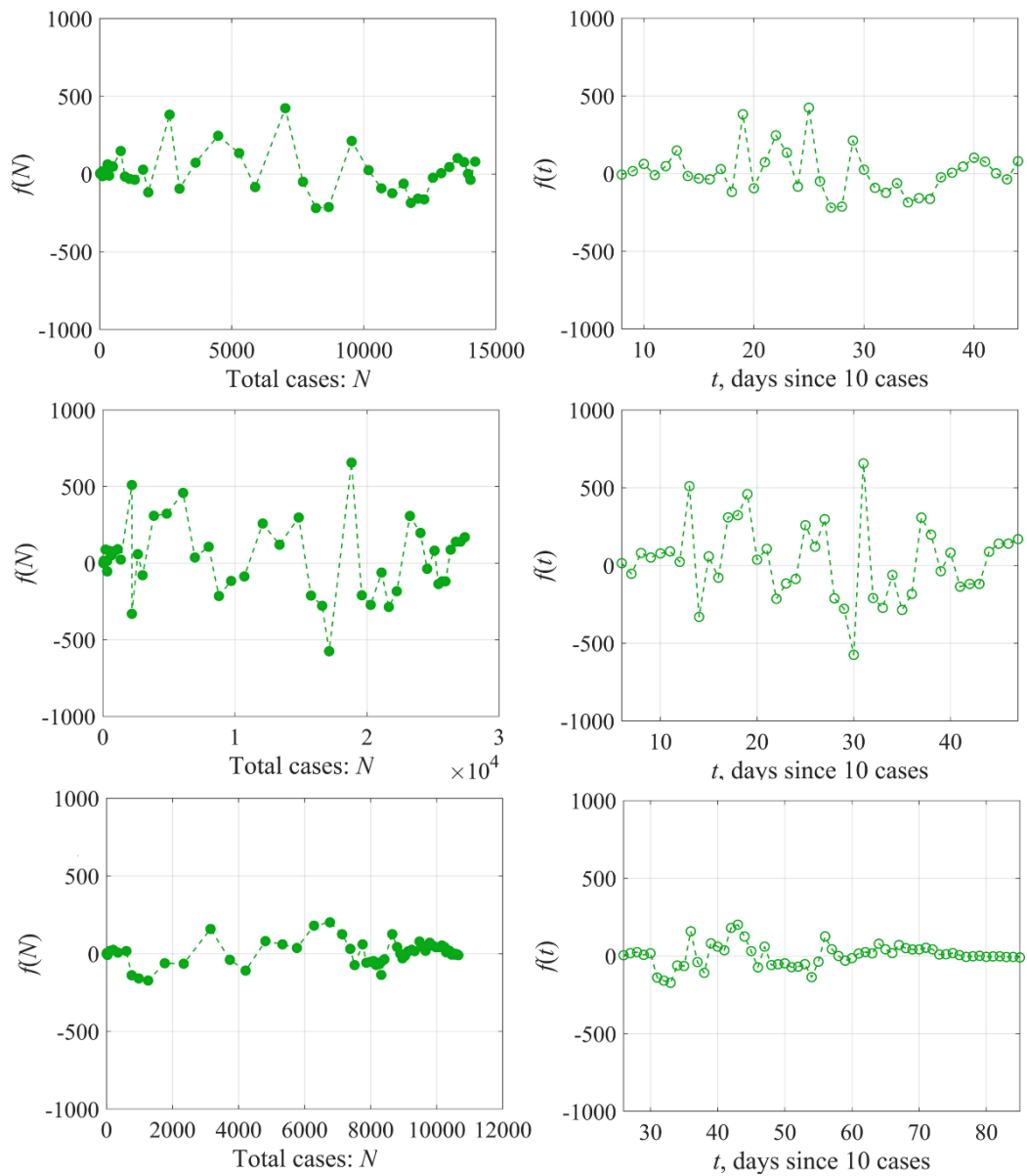

Fig. 11. «External force» $f$ as a function of the number of cases $N$ (left) and $f(t)$ (right) in the framework of the generalized logistic model for Austria, Switzerland and South Korea (top to bottom).

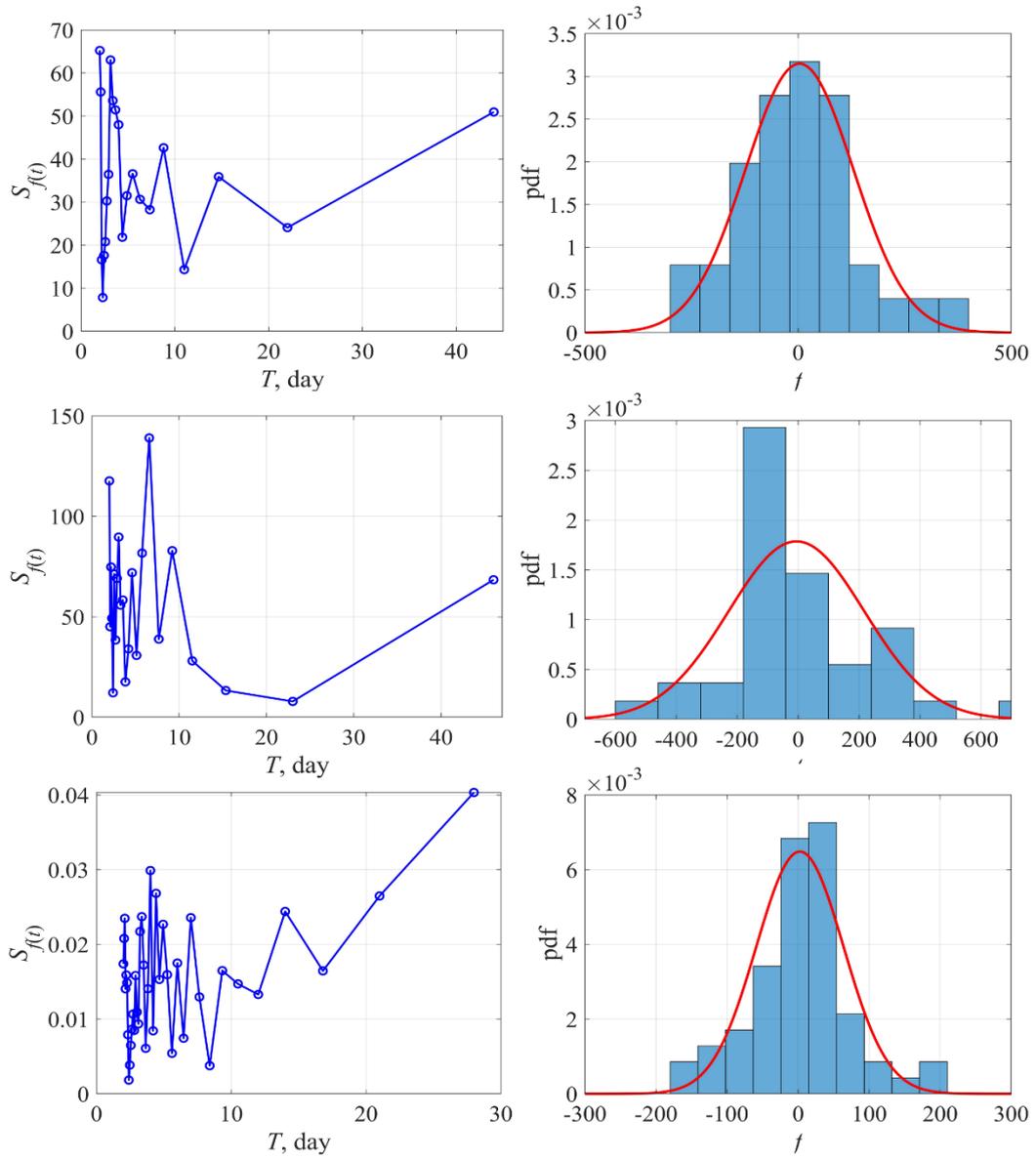

Fig. 12. Amplitude spectrum of *f(t)* – the left column, histogram and its Gaussian approximation (the red line) – the right column, for Austria, Switzerland and South Korea (top to bottom). Distribution parameters of *f(t)* for Austria: standard deviation 125, skewness 0.6, kurtosis 1.5; Switzerland: standard deviation 220, skewness 0.9, kurtosis 2.6; South Korea: standard deviation 61, skewness 0.2, kurtosis 2.5.

Thus, the generalized logistic equation can be considered to be a stochastic one with time-dependent coefficients

$$\frac{dN}{dt} = r(t)N^{\alpha}\left[1 - p(t)N\right]^{\beta} + f(t), \qquad (19)$$

or with non-linear functions depending on the number of cases *N*.

## 4. Discussion and conclusions

Summarizing the results, we would like to emphasize, that with all its simplicity and crudity, the logistic model describes properly the growth in the number of COVID-19 cases with time. This is illustrated by Fig. 13, which shows the actual data and logistic curves. It is evident that for many countries the use of the simple logistic equation leads to a very good agreement with the available data. The use of a generalized logistic curve improves the agreement significantly, including the countries for which the logistic model is too crude (for example, South Korea). It is worth noting that the prognostic number of the total number of cases $N_\infty$ in the generalized logistic model is slightly higher than in the simple logistic model, and the approach to the limiting constant value is delayed in time.

Fig. 14 illustrates the capabilities of the logistic model for describing the number of sick people per day. On average, the theoretical model describes the real data rather well, but the scatter of points is still not small, and sometimes deviations can reach 50% and higher, although on average they are less than 50%. These differences are especially evident near the epidemic peak when it is desirable to have a more accurate prognosis for medical facilities. The extent to which this scatter is better described by other models (such as SIR models) will be clear in the near future when the results of relevant studies appear. From the mathematical point of view, the resulting difference in the use of the logistic model to describe two characteristics: the total number of cases (*N*) and the number of cases per day (*K*) is obvious: the curve $N(t)$ is the integral with respect to $K(t)$, and, therefore, it is smoother and more determined. To describe the dependence $K(t)$, at least on a qualitative level, it is better to use stochastic equations of a logistic model of the Eq. (19) type, where external random forces or random coefficients are introduced. They will help to understand the degree of the data spread, and, what is most important, the number of possible large outliers during the epidemic. Such work remains to be done.

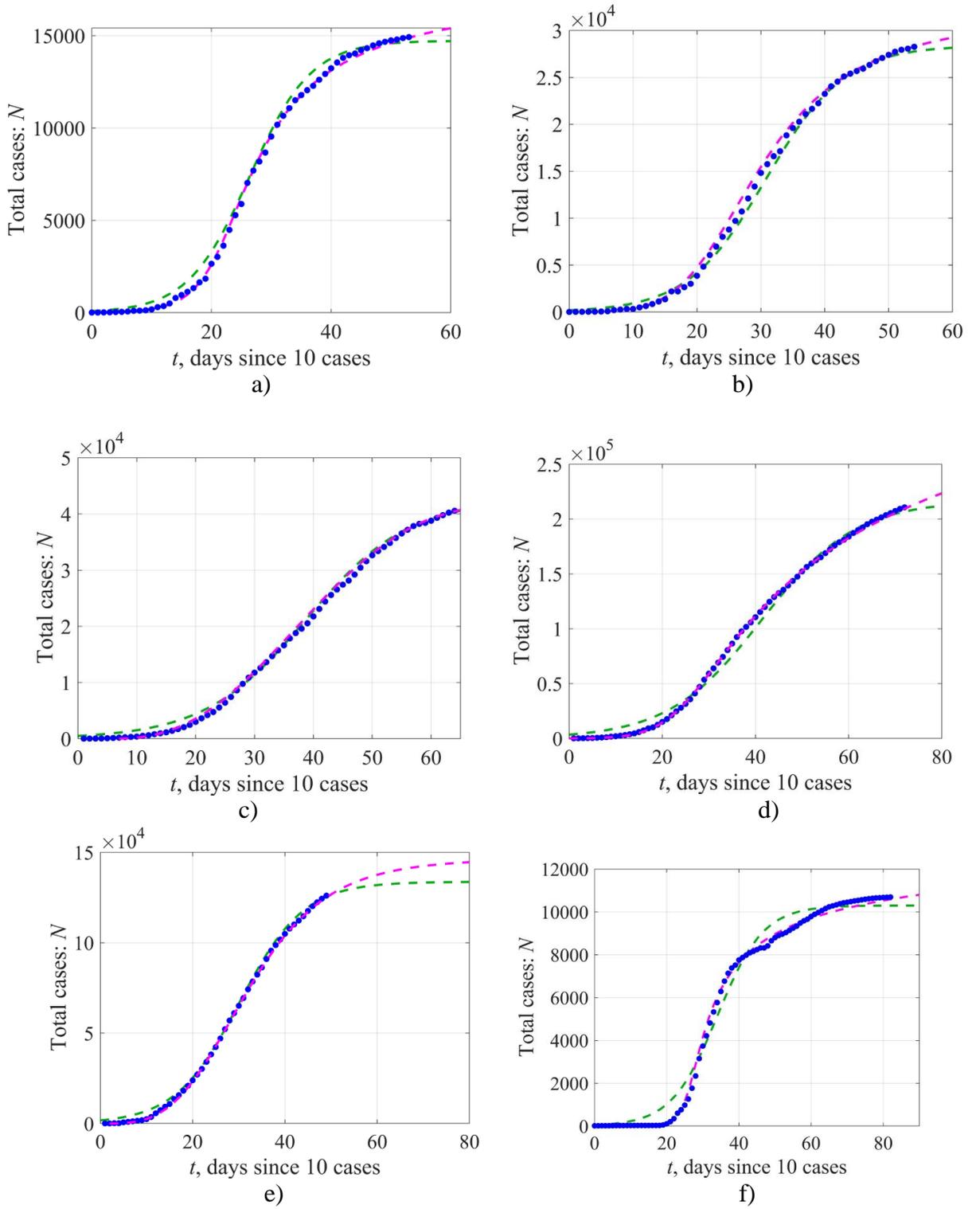

Fig. 13. The total number of cases in time: blue markers represent the initial data, the green dashed line – solution (2) of the simple logistic model, the pink dashed line – the numerical solution of Eq. (14) of the generalized logistic model for Austria (*a*), Switzerland (*b*), the Netherlands (*c*), Italy (*d*), Turkey (*e*) and South Korea (*f*) (top to bottom).

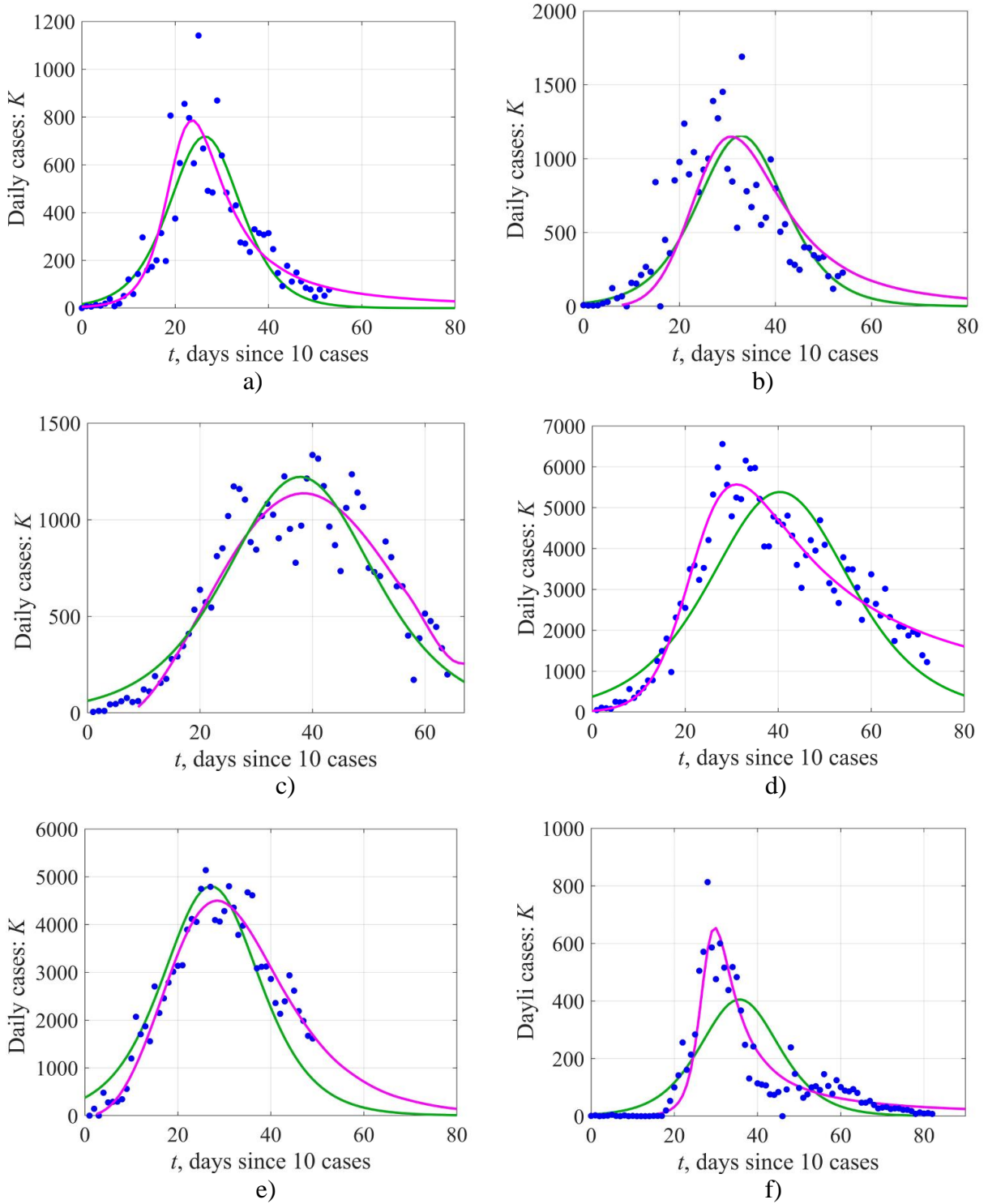

Fig. 14. The number of cases per day in time: blue markers represent the initial data, the green line corresponds to expression (4) in the framework of the logistic equation, Eq. (1), of the simple logistic model, the pink line – numerical solution of Eq. (14) of the generalized logistic model for Austria (*a*), Switzerland (*b*), the Netherlands (*c*), Italy (*d*), Turkey (*e*) and South Korea (*f*).

The authors understand that for the real forecast of the epidemic development, it is necessary to have multifactor models, which include dividing the population into different groups (children, the elderly, etc.), living conditions (traffic flows between territories, the population density etc.). Such models should include high-order ODEs and PDEs, taking into account lagging arguments and integral terms. Such complex models are being developed now, yet we will not consider them here. Nevertheless, the analysis within the framework of simple low-parameter models is important because it allows us to describe the process qualitatively, to understand the role of certain factors, and to identify certain phenomena (stochastization, fractality, nonlinearity) that are also interesting for other branches of physics and technology. In this sense, the results obtained above demonstrate the capabilities of a well-developed logistic model for describing an epidemic of such a grand scale as COVID-19.

**Acknowledgements**

The presented results were obtained with the financial support of the grant of the President of the Russian Federation for state support of scientific research of leading scientific schools of the Russian Federation NSh-2485.2020.5. EP is partially supported by the Laboratory of Dynamical Systems and Applications NRU HSE, of the Ministry of science and higher education of the RF, grant ag. № 075-15-2019-1931.

[42]   **Zhang, Xiaolei, Renjun Ma, and Lin Wang.** "Predicting turning point, duration and attack rate of COVID-19 outbreaks in major Western countries." *Chaos, Solitons & Fractals* 135 (2020): 109829. doi: doi.org/10.1016/j.chaos.2020.109829

[43]   **Matabuena, Marcos, et al.** "COVID-19: Estimating spread in Spain solving an inverse problem with a probabilistic model." *arXiv preprint arXiv:2004.13695* (2020).

[44]   **Heinsalu, Els, David Navidad Maeso, and Marco Patriarca.** "The dynamics of natural selection in dispersal-structured populations." *Physica A: Statistical Mechanics and its Applications* 547.1 (2020): 124427. doi: 10.1016/j.physa.2020.124427

[45]   **Li, Lixiang, et al.** "Propagation analysis and prediction of the COVID-19." *Infectious Disease Modelling* 5 (2020): 282-292. doi: 10.1016/j.idm.2020.03.002
31